\documentclass[10pt]{article}
\usepackage{fullpage}
\usepackage{amsmath}
\usepackage{amsfonts}
\newcommand{\tr}{\mathop{\rm tr}\nolimits}
\newcommand{\sgn}{\mathop{\rm sign}\nolimits}
\newcommand{\arccosh}{\mathop{\rm arccosh}\nolimits}
\begin{document}
\title{ \sc{Two-dimensional electron liquid with disorder in a weak
magnetic field} }
\author{I.S. Burmistrov \thanks{E-mail: burmi@itp.ac.ru} \\
{\small L.D. Landau Institute for Theoretical Physics, Kosygina
str. 2, 117940 Moscow, Russia }}
\date{}
\maketitle
\begin{flushleft}PACS: \qquad 73.43 -f \qquad 73.20.Mf
\qquad 73.43.Jn \end{flushleft}

\begin{abstract}
\par We present the effective theory for low energy dynamics of a
two-dimensional interacting electrons in the presence of a weak
short-range disorder and a weak perpendicular magnetic field, the
filling factor $\nu \gg 1$. We investigate the exchange
enhancement of the $g$-factor, the effective mass and the decay
rate of the simplest spin wave excitations at $\nu = 2 N + 1$. We
obtain the enhancement of the field-induced gap in the tunneling
density of states and dependence of the tunneling conductivity on
the applied bias.
\end{abstract}


\section{Introduction}
\label{Intro}

\par A two-dimensional electron gas in the perpendicular magnetic field
has attracted much attention both from the theoretical and from
the experimental points of view. The effects in a strong magnetic
field when only the lowest Landau level is occupied have been
investigated since the discovery of the quantum Hall
effect~\cite{QHE}. Several efforts~\cite{old} are made in order to
involve the larger filling factors $\nu > 1$ into the problem
discussed. However, an existence of the small parameter, namely, a
 ratio of the Coulomb energy at the magnetic field length to the
cyclotron energy has been assumed. In fact, in a weak magnetic
field the Coulomb energy at the magnetic field length exceeds the
cyclotron energy and some attempts~\cite{SMG} have been undertaken
to investigate the case of the large filling factor $\nu > 1$.
\par The experimental investigation of the tunneling density of
states for the system considered was performed at small ($\nu <
1$)~\cite{EPW} and at large ($\nu > 1$)~\cite{ALBS} filling
factors. In the case of a weak magnetic field ($\nu \gg 1$) the
gap in the tunneling density of states has been obtained in the
framework of hydrodynamical approach~\cite{ABG}. The progress was
made by Aleiner and Glazman~\cite{AG} who have developed the
effective theory for low-energy excitations on the partially
filled Landau level at a large filling factors $\nu \gg 1$.
\par Recently, after the prediction that the unidirectional
charge-density wave state occurs at half-filled high Landau levels
within the framework of Hartree-Fock theory~\cite{HF} and the
experimental discovery of compressible states with the anisotropic
 magnetotransport properties in the high mobility systems near the
 half-filling of the high Landau levels~\cite{Exp},
  the two-dimensional electron liquid in a weak magnetic field
  has been studying intensively~\cite{Fogler}.
\par In the paper the low energy effective theory for electrons at
the partially filled Landau level with the  large filling factor
in the presence of disorder is developed (Sec.\ref{EffAct}). As an
example, the effect of disorder on the exchange enhancement of the
effective $g$-factor and the simplest spin-wave excitations are
discussed in Sec.\ref{Spin}. An electron tunneling into the
electron liquid is considered in Sec.\ref{zba}. The conclusion is
given in Sec.\ref{Conc}.

\section{Derivation of the effective action}
\label{EffAct}

\subsection{Introduction}
\label{Intro.EffAct}

\par We consider the system of two-dimensional electrons with the Coulomb
interaction in the presence of disorder and a perpendicular
magnetic field $H$. The system possesses a partially filled high
Landau level with the level index $N \gg 1$ equal to the integer
part of a half of the filling factor $\nu$, namely, $N = [\nu/2]$.
The presence of a random potential which is considered to be
short-range results in a broadening of the Landau levels. We imply
that the elastic collision time satisfies the condition $\tau_{0}
\gg \omega_{c}^{-1}$ where $\omega_{c} = e H / m$ is the cyclotron
frequency with the charge of electrons $e$ and the electron mass
$m$. In this case the broadening of the Landau levels which is of
order of $\sqrt{\omega_{c} \tau_{0}} / \tau_{0}$ is much less then
the distance between them.
\par The conventional parameter characterizing the coupling strength of
the Coulomb interaction is $r_{s} = \sqrt{2} e^{2}/v_{F}$ with
$v_{F}$ being the Fermi velocity. We assume that electrons are
weakly interacting, i.e., $r_{s} < 1$. In this case we can treat
the problem discussed in the random phase approximation. We also
assume that the number $N$ is sufficiently large, so the condition
$N r_{s} \gg 1$ satisfies. This means that the cyclotron radius
$R_{c} = \sqrt{\nu / m \omega_{c}}$ is supposed to be much more
than the Bohr radius $a_{B} = 1/m e^{2}$, namely, $R_{c} \gg
a_{B}$.

\subsection{Formalism}
\label{EffAct.Form}

The system is described by the following grand canonical partition
function in the path-integral representation
\begin{equation}
Z = \int {\cal D}[ \overline{\psi}, \psi] \int {\cal D} [V_{dis}]
\,{\cal P}[V_{dis}(\vec r)]\, e^{\displaystyle S[ \overline{\psi},
\psi, V_{dis}]} \label{ZStart}
\end{equation}
\begin{eqnarray}
 S  & = & \sum \limits_{\alpha=1}^{N_{r}} \int \limits_{0}^{1/T} d \tau
 \int d^{2} { \vec r}\, \Bigl [ \overline{\psi}^{\alpha,\sigma}(\vec
r,\tau) \left ( -\partial_{\tau} + \mu - {\cal H}
 - V_{dis}(\vec r) \right ) \psi^{\alpha,\sigma}(\vec
r,\tau) \nonumber \\ & - &  \frac{1}{2}\int d^{2} { \vec r_{1}}
\,\overline{\psi}^{\alpha,\sigma}(\vec r,\tau)
\psi^{\alpha,\sigma}(\vec r,\tau) U_{0}(\vec r,\vec r_{1})
\overline{\psi}^{\alpha,\sigma_{1}}(\vec r_{1},\tau)
\psi^{\alpha,\sigma_{1}}(\vec r_{1},\tau) \Bigr ] \label{ActStart}
\end{eqnarray}
Here $ \psi^{\alpha,\sigma}$ and $\overline{\psi}^{\alpha,\sigma}$
are the Grassmann variables defined within the imaginary time
interval $ \tau \in [0,1/T]$ with the antiperiodic condition
$\psi(\vec r,1/T) = - \psi(\vec r,0) $. Symbol $T$ stands for  the
temperature, $\mu$ is the chemical potential of the system, and
$\sigma,\sigma_{1} = \pm 1$ are spin indices. Hamiltonian ${\cal
H} = ( -i \nabla - e \vec A)^{2} / (2 m)$ describes an electron
with mass $m$ propagating in the two-dimensional space in the
perpendicular magnetic field $H = \epsilon_{a b}
\partial_{a} A_{b}$. The random potential $V_{dis}(\vec r)$  is
chosen to have the Gaussian distribution function
\begin{equation} {\cal P}[V_{dis}(\vec r)] = \sqrt{\rho
\tau_{0}}\, \exp \left ( - \pi \rho \tau_{0} \int d^{2} { \vec r}
\,V_{dis}^{2}(\vec r) \right ) \label{dis}
\end{equation}
where $\rho$ denotes the thermodynamical density of states.
\par In order to find the
average $\overline{\ln Z}$ over disorder, there are introduced
$N_{r}$ replicated copies of the system and $\alpha=1,...,N_{r} $
are replica indices.
\par The Matsubara representation seems to be more convenient for the
problem concerned. Therefore the Fourier transform from imaginary
time $\tau$ to the Matsubara frequencies will be employed. Since
the fermionic fields are antiperiodic within the interval
$[0,1/T]$, the frequencies permitted for $\overline{\psi}$ and
$\psi$ are $ \omega_{n} = \pi T ( 2 n + 1 ) $, $n$ being an
integer. The Fourier-transformed fields are defined, respectively
\begin{equation}
\begin{array}{lcr}
\overline{\psi}^{\alpha}(\tau) = \sqrt{T} \sum
\limits_{n=-\infty}^{\infty} \overline{\psi}^{\alpha}_{n} e^{i
\omega_{n} \tau} & , & \psi^{\alpha}(\tau) = \sqrt{T} \sum
\limits_{n=-\infty}^{\infty} \psi^{\alpha}_{n} e^{-i \omega_{n}
\tau}
\end{array}
\end{equation}
Later on, the limits in the frequency and replica series will be
omitted for a brevity.
\par In the Matsubara representation the action (\ref{ActStart}) becomes
\begin{eqnarray}
 S & = &
 \int d^{2} { \vec r}\,  \sum \limits_{\alpha,n}
\Bigl [ \overline{\psi}^{\alpha,\sigma}_{n}(\vec r) \left ( i
\omega_{n} + \mu - {\cal H} - V_{dis}(\vec r) \right )
\psi^{\alpha,\sigma}_{n}(\vec r)
 \nonumber \\
 & -  &  \frac{T}{2}\sum \limits_{l,m} \int
d^{2} { \vec r_{1}}\,  \overline{\psi}^{\alpha,\sigma}_{m}(\vec r)
\psi^{\alpha,\sigma}_{m-n}(\vec r) U_{0}(\vec r,\vec r_{1})
\overline{\psi}^{\alpha,\sigma_{1}}_{l}(\vec r_{1})
\psi^{\alpha,\sigma_{1}}_{l+n}(\vec r_{1}) \Bigr ] \label{ActM}
\end{eqnarray}
\par The Zeeman term in action (\ref{ActStart}) is neglected
due to smallness of the $g$-factor. In fact, the condition $g \ll
1$ is usually hold. Nevertheless, the Zeeman term can be involved
into the effective action after performing the integration over
the fast degrees of freedom. To simplify the notation, the spin
indices will be associated with the replica ones where it will be
convenient.

\subsection{Plasmon field and the average over disorder}
\label{EffAct.PandD}

The Coulomb term in action (\ref{ActM}) is quartic in the
fermionic fields. One can get rid of this quartic term by
performing the Hubbard-Stratonovich transformation, introducing an
extra path integration over bosonic field
$\lambda^{\alpha}_{n}(\vec r)$. With the help of the so-called
plasmon field the Coulomb term can be presented as
\begin{eqnarray}
\int {\cal D}[\lambda] \exp \left [ -\frac{T}{2} \int\!\!\int
d^{2} { \vec r}\, d^{2} { \vec r_{1}}\, \lambda^{\dagger}(\vec r)
U_{0}^{-1}(\vec r,\vec r_{1}) \lambda(\vec r_{1}) + i T \int d^{2}
{ \vec r}\, \psi^{\dagger}(\vec r) \hat \lambda(\vec r) \psi(\vec
r) \right ] \label{HSL}
\end{eqnarray}
where $U_{0}^{-1}$ stands for the operator inverse in $U_{0}$. The
 matrix notations are used for the combined replica and frequency
indices
\begin{equation}
\begin{array}{lcr}
\psi^{\dagger}(\cdots)\psi = \sum \limits_{n, m}^{\alpha,\beta}
{\overline \psi}^{\alpha}_{n} (\cdots)^{\alpha \beta}_{n m}
\psi^{\beta}_{m} & , & \lambda^{\dagger}\lambda = \sum
\limits_{n}^{\alpha} \lambda^{\alpha}_{-n} \lambda^{\alpha}_{n}
\end{array}
\end{equation}
The quantities with the ``hat'' are defined according to $\hat z =
\sum_{\alpha,n} z^{\alpha}_{n} I^{\alpha}_{n}$ with the matrix
$(I^{\alpha}_{n})^{\beta \gamma}_{k l} = \delta^{\alpha \beta}
\delta^{\alpha \gamma} \delta_{k-l,n}$. The matrices
$I^{\alpha}_{n}$ represent the diagonals shifted in the frequency
space and, in  general, are the generators of the $U(1)$-gauge
transformations. The measure of the functional integral over the
plasmon field $\lambda$ is introduced so that integral (\ref{HSL})
equals unity while fermionic fields $\psi^{\dagger}$ and $\psi$
vanish.
\par In order to perform the averaging over disorder in the partition function
(\ref{ZStart}), one should integrate over the random potential
$V_{dis}(\vec r)$. This leads straightforwardly to the following
quartic term
\begin{equation}
\frac{1}{4 \pi \rho \tau} \int d^{2} { \vec r}\, \sum \limits_{n
m}^{\alpha \beta} \overline \psi^{\alpha}_{n}(\vec r)
\psi^{\alpha}_{n}(\vec r) \overline \psi^{\beta}_{m}(\vec r)
\psi^{\beta}_{m}(\vec r) \label{quartic2}
\end{equation}
in the action. The term (\ref{quartic2}) can be decoupled by means
of the Hubbard-Stratonovich transformation. An extra path
integration over the Hermitian matrix field variables $Q^{\alpha
\beta}_{n m}(\vec r)$ can be introduced~\cite{ELK,Finkel}
\begin{equation}
\int {\cal D}[Q] \exp \int d^{2} { \vec r}\, \left [  - \pi \rho
\tau_{0} \tr Q^2(\vec r)   + i \psi^{\dagger}(\vec r) Q(\vec r)
\psi(\vec r)\right ]  \label{HSQ}
\end{equation}
Here the symbol $\tr$ denotes the matrix trace over the Matsubara,
replica and spin spaces. The measure of the functional integral
over the matrix field $Q$ is defined in the same way as for the
plasmon field, i.e., integral (\ref{HSQ}) equals unity while
fermionic fields $\psi^{\dagger}$ and $\psi$ vanish.
\par After calculations discussed above the partition function becomes
\begin{equation}
Z = \int {\cal D}[\overline{\psi},\psi,\lambda,Q]\,
e^{\displaystyle S[\overline{\psi},\psi,\lambda,Q]} \label{Znext}
\end{equation}
\begin{eqnarray}
S & = & -  \pi \rho \tau_{0} \int d^{2} { \vec r}\, \tr Q^2
 - \frac{T}{2} \int\!\!\int d^{2} { \vec r}\, d^{2} { \vec r_{1}}\, \lambda^{\dagger}(\vec r) U_{0}^{-1}(\vec r,\vec r_{1})
\lambda(\vec r_{1}) \nonumber \\ & + & \int d^{2} { \vec r}\,
 \psi^{\dagger}(\vec r) \left ( i \omega + \mu -
{\hat {\cal H}} + i T \hat \lambda + i Q \right ) \psi(\vec r)
\label{S2}
\end{eqnarray}
Here $\omega$ is a unity matrix in the replica space while in the
Matsubara space it is a matrix containing the frequencies
$\omega_{n}$ on the diagonal, namely, $(\omega)^{\alpha \beta}_{n
m} = \omega_{n} \delta_{n m} \delta^{\alpha \beta} $.

\subsection{Elimination of the $N$-th Landau level}
\label{EffAct.Exclus}

The fermionic fields $\psi^{\dagger}$ and $\psi$ refer to all
Landau levels. In order to integrate over all the fermionic
degrees of freedom not belonging to the partially filled $N$-th
Landau level, we separate the fermionic fields into two kinds. The
first field refers to the $N$-th Landau level
\begin{equation}
\begin{array}{lcr}
\Psi(\vec r) = \sum \limits_{k} \psi_{N k} \varphi_{N k}(\vec r) &
, &  \Psi^{\dagger}(\vec r) = \sum \limits_{k} \psi^{\dagger}_{N
k} \varphi_{N k}(\vec r)
\end{array}
\label{Psi}
\end{equation}
The second one involves the other levels
\begin{equation}
\begin{array}{lcr}
\Phi(\vec r) = \sum \limits_{p\neq N,k} \psi_{p k} \varphi_{p
k}(\vec r) & , &  \Phi^{\dagger}(\vec r) = \sum \limits_{p \neq N,
k} \psi^{\dagger}_{p k} \varphi_{p k}(\vec r)
\end{array}
\label{phi}
\end{equation}
where $\varphi_{p k}(\vec r)$ are the eigenfunctions of the
hamiltonian $\cal H$ and $p=0,1,...,N,...$ numerates the Landau
levels with energies $\epsilon_{p} = \omega_{c} ( p+ 1/2) $. In
addition, we introduce two types of the Green functions. One is
for the $N$-th Landau level
\begin{equation}
G(\vec r,\vec r_{1}; Q, \lambda) = \sum \limits_{k, k^{\prime}}
\varphi^{*}_{N k}(\vec r) G_{N k, N k^{\prime}}(Q, \lambda)
\varphi_{N k^{\prime}}(\vec r_{1}) \label{GreenN}
\end{equation}
and another is for the other levels
\begin{equation}
{\tilde G}(\vec r,\vec r_{1}; Q, \lambda) = \sum \limits_{p,
p^{\prime} \neq N} \sum \limits_{k, k^{\prime}} \varphi^{*}_{p
k}(\vec r) G_{p k, p^{\prime} k^{\prime}}(Q, \lambda)
\varphi_{p^{\prime} k^{\prime}}(\vec r_{1}) \label{Green}
\end{equation}
where the inverse of the Green function for the $\psi_{p k}$ and
$\psi^{\dagger}_{p^{\prime} k^{\prime}}$ operators is as follows
\begin{equation}
(G^{-1})_{p k, p^{\prime} k^{\prime}} = ( i \omega + \mu -
\epsilon_{p}) \delta_{p p^{\prime}} \delta_{k k^{\prime}} + i T
{\hat \lambda}_{p k, p^{\prime} k^{\prime}} + i Q_{p k, p^{\prime}
k^{\prime}} \label{GreenInv}
\end{equation}
with the following matrix elements
\begin{equation}
f_{p k, p^{\prime} k^{\prime}} = \int d^{2} \vec r \,
\varphi^{*}_{p^{\prime} k^{\prime}}(\vec r) f(\vec r) \varphi_{p
k}(\vec r)
\end{equation}
\par The action (\ref{S2}) is bilinear in the fermionic fields $\psi^{\dagger}$
and $\psi$ and obviously does in the fermionic fields
$\Phi^{\dagger}$ and $\Phi$ too. Therefore one can integrate over
the fermionic fields $\Phi^{\dagger}$ and $\Phi$ and obtain the
following result
\begin{eqnarray}
& & S  =  - \int \tr \ln \tilde G - \pi \rho \tau_{0} \int \tr Q^2
+ \int
 \Psi^{\dagger} \left [ i \omega + \mu -
{\hat {\cal H}} + i T \hat \lambda + i Q \right ] \Psi  \nonumber
\\ & - &
  \frac{T}{2} \int\!\!\int \lambda^{\dagger} U_{0}^{-1}
\lambda  +   \int\!\!\int \left [ \Psi^{\dagger} Q {\tilde G} Q
\Psi +  2 T  \Psi^{\dagger} \hat \lambda {\tilde G} Q \Psi   +
T^{2} \Psi^{\dagger} \hat \lambda {\tilde G} \hat \lambda \Psi
\right ] \label{S3}
\end{eqnarray}
Hereafter, the space indices are omitted. It should be noted that
the last term in the action (\ref{S3}) appears due to the
interaction between electrons belonging to the partially filled
$N$-th Landau level and the other electrons.

\subsection{ Integration over the $Q$ field}
\label{EffAct.Q}

The $Q$ matrix field should be divided  into the transverse $V $
and the longitudinal $ P $ components as follows $Q = V^{-1} P V$.
Here the longitudinal component $P$ has a block-diagonal structure
in the Matsubara space, i.e., $P^{\alpha \beta}_{n m} \propto
\Theta(n m)$ where $\Theta(x)$ is the Heaviside step function. The
 transverse component $V$ corresponds to a unitary rotation. One can find a
 review on the above discussion in Refs.~\cite{P1,
BPS1}.
\par The change of variables $Q$ with $P$ and $V$ is motivated
by the saddle-point structure of the action (\ref{S3}) in the
absence of the plasmon field $\lambda$ and at zero temperature,
i.e., $\omega_{n} \to 0$. This saddle-point solution can be
written as  $Q_{sp} = V^{-1} P_{sp} V$ where the matrix $P_{sp}$
obeys the equation
\begin{equation}
2 \pi \rho \tau_{0} P_{sp} = i \Bigl [ G_{0}(\vec r, \vec r) +
{\tilde G}_{0}(\vec r, \vec r) \Bigr ] \label{sp}
\end{equation}
which coincides with the self-consistent Born approximation
equation~\cite{Ando}. Here the Green function $ G_{0}$ is
determined as a special case of $G$, namely, $ G_{0}(\vec r, \vec
r_{1}) = G(\vec r, \vec r_{1}; P_{sp}, 0) $ and the same for $
\tilde G_{0}$.
\par In the case of small disorder $\omega_{c} \tau_{0} \gg 1 $ the
solution of equation (\ref{sp}) has the form as
\begin{equation}
\begin{array}{lcr}
\displaystyle  (P_{sp})_{n m}^{\alpha \beta} = \frac{\sgn n}{2
\tau} \delta_{n m} \delta^{\alpha \beta} & , & \displaystyle \tau
= \pi \sqrt{\frac{\rho}{m}} \frac{\tau_{0}}{\sqrt{\omega_{c}
\tau_{0}}}
\end{array}
\label{sp1}
\end{equation}
\par The presence of the plasmon field $\lambda$ results in a
shift of the saddle-point value (\ref{sp1}) of the $P$ field which
can be found by expanding the action (\ref{S3}) to second order in
both $\lambda$ and $\delta P = P - P_{sp}$. Hence one can obtain
\begin{equation}
S  =  S_{0} + S_{1}[\delta P, \lambda] + S_{2}[\overline \Psi,
\Psi, \delta P, \lambda] \label{S4}
\end{equation}
\begin{equation}
 S_{0}  =  \int  \left (
\tr \ln {\tilde G}_{0} - \pi \rho \tau_{0} \tr Q_{sp}^{2}  +
\Psi^{\dagger} \left [ i \omega + \mu - {\hat {\cal H}} +
 i T \hat \lambda + i  Q_{sp} \right ] \Psi \right )
\label{S0}
\end{equation}
\begin{equation}
S_{1}  =  i T \int \tr {\tilde G}_{0} \hat \lambda - \pi \rho
\tau_{0} \int \tr (\delta P)^{2} - \frac{T}{2} \int\!\!\int
\lambda^{\dagger} U_{0}^{-1} \lambda -
 \frac{1}{2} \int\!\!\int \tr \left [ T \hat \lambda + \delta P
\right ]
 \pi_{0}
\left [ T \hat \lambda + \delta P \right ] \label{Ss1}
\end{equation}
\begin{equation}
S_{2}  =  \int\!\!\int \left ( \Psi^{\dagger} \left [ T \hat
\lambda + \delta P \right ] \tilde G_{0} \left [ T \hat \lambda +
\delta P \right ]\Psi
 - 2 \tr
\left [ T \hat \lambda + \delta P \right ] \tilde G_{0} \left [ T
\hat \lambda + \delta P \right ] G_{0}\right ) \label{Ss2}
\end{equation}
where a bare polarization operator $\pi_{0}$ is implied to be a
matrix according to the rule
\begin{eqnarray}
\tr A \pi_{0} B = \sum \limits_{n m}^{\alpha \beta}
A_{m+n,m}^{\alpha \beta}(\vec r) \pi_{0}^{m}(n;\vec r,\vec r_{1})
B_{m,m+n}^{\beta \alpha }(\vec r_{1})
\end{eqnarray}
and is defined by
\begin{equation}
\pi^{m}_{0}(n; \vec r, \vec r_{1})  =  - 2 \left ( \tilde
G_{0}^{m+n}(\vec r, \vec r_{1}) \tilde G_{0}^{m}(\vec r_{1}, \vec
r)
+\tilde G_{0}^{m+n}(\vec r, \vec r_{1}) G_{0}^{m}(\vec r_{1}, \vec
r) + G_{0}^{m+n}(\vec r, \vec r_{1}) \tilde G_{0}^{m}(\vec r_{1},
\vec r) \right )
\end{equation}
\par After decomposing of the matrix field $Q$ into the block-diagonal Hermitian
matrix field $P$ and unitary matrix field $V$, the measure of the
functional integral in (\ref{S4}) becomes ${\cal D}[Q] = {\cal
D}[V]{\cal D}[\delta P] I[\delta P]$ where~\cite{P1}
\begin{equation}
\ln I[\delta P] = -\frac{1}{(\pi \rho)^{2}}\int \sum \limits_{n
m}^{\alpha \beta} \left [ 1 - \Theta(n m) \right ] \delta P_{n
n}^{\alpha \alpha} \delta P_{m m}^{\beta \beta} \label{mesI}
\end{equation}
\par The terms which are quadratic in $\delta P$ from the term $S_{1}$ of action
(\ref{S4}) together with the contribution from measure
(\ref{mesI}) determine the propagator of the $\delta P$ fields
\begin{eqnarray}
 \langle \delta
P_{m_{1} m_{2}}^{\alpha \beta}(\vec q) \delta P_{m_{3}
m_{4}}^{\gamma \delta}(-\vec q) \rangle & = &  \frac{
\displaystyle \delta_{m_{1} m_{4}} \delta_{m_{2} m_{3}}
\delta^{\alpha \delta} \delta^{\beta \gamma} \frac{\Theta(m_{1}
m_{3})}{2 \pi \rho \tau_{0}}} {\displaystyle 1 +
\frac{\pi_{0}^{m_{1}}(m_{3}-m_{1};\vec q)}{2 \pi \rho
\tau_{0}}}  \nonumber \\
& - &  \frac{2 \left [ 1 - \Theta(m_{1} m_{3}) \right ]}{(2
\pi^{2} \rho^{2} \tau)^{2}} \frac{ \delta_{m_{1}
m_{2}}\delta^{\alpha \beta}}
 {\displaystyle 1 + \frac{\pi_{0}^{m_{1}}(0;\vec q)}{2 \pi \rho
\tau_{0}} } \frac{\delta_{m_{3} m_{4}} \delta^{\delta \gamma}}
 {\displaystyle 1 + \frac{\pi_{0}^{m_{3}}(0;\vec q)}{2 \pi \rho
\tau_{0}} }  \label{Pcor}
\end{eqnarray}
It should be noted that the propagator of the longitudinal
fluctuations (\ref{Pcor}) proves to be analogous to that
previously obtained for the problem on the behavior of a free
electron gas in the perpendicular magnetic field~\cite{P1}.
\par Using the expression (\ref{Pcor}) for the propagator of the $\delta P$
fields, one can integrate action (\ref{S4}) over the longitudinal
fluctuations in the quadratic approximation. This yields the
following result
\begin{equation}
S = S_{0} + S_{\lambda} + S_{\mu}
\label{S5}
\end{equation}
where $S_{0}$ given by equation (\ref{S0}) describes the electrons
at the partially filled $N$-th Landau level coupled to the plasmon
and $Q_{sp}$ fields. The term $S_{\lambda}$ concerns the screening
of the Coulomb interaction due to the influence of electrons from
the other Landau levels and is given by
\begin{eqnarray}
S_{\lambda}  =  i T \int d^{2} \vec r \, \tr {\tilde G}_{0}(\vec
r, \vec r) \hat \lambda(\vec r)
-\frac{T}{2} \int \frac{d^{2} { \vec q}}{(2 \pi)^{2}} \sum
\limits_{n}^{\alpha} \lambda^{\alpha}_{-n}(\vec q) U_{0}^{-1}(q)
\varepsilon(n,\vec q) \lambda^{\alpha}_{n}(- \vec q)
\label{Slambda}
\end{eqnarray} where the
dielectric function $ \varepsilon(n,q) = 1 + U_{0}(q) \Pi(n,q) $
with the following polarization operator~\footnote{ It should be
noted that similar form of the polarization operator but with
another bare polarization operator $\pi_{0}^{m}(n,q)$ was first
derived by Baranov and Pruisken~\cite{BPun}}
\begin{eqnarray}
\Pi(n,q) & = & T \sum \limits_{m} \pi_{0}^{m}(n,q) \Biggl [ 1 -
 \frac{ \displaystyle \Theta(n(n+m)) \frac{\pi_{0}^{m}(n,q)}{2 \pi \rho
 \tau_{0}}
}{\displaystyle 1+ \frac{\pi_{0}^{m}(n,q)}{2 \pi \rho \tau_{0}} }
\Biggr ] \nonumber
\\ & + & T \frac{\delta_{n,0}}{(\pi^{2} \rho^{2} \tau)^{2}} \sum \limits_{k m}
\frac{ [1 - \Theta(k m)] \pi_{0}^{m}(0,q)}{\displaystyle 1+
\frac{\pi_{0}^{m}(0,q)}{2 \pi \rho \tau_{0}}} \frac{
\pi_{0}^{k}(0,q)}{\displaystyle 1+ \frac{\pi_{0}^{k}(0,q)}{2 \pi
\rho \tau_{0}} } \label{Pi}
\end{eqnarray}
The third term $S_{\mu}$ of action (\ref{S5}) contains the terms
which affect the chemical and thermodynamic potentials of the
system (See Appendix A).
\par It is worthwhile to mention that the saddle-point
approximation in which the integration over the $Q$ field is
performed is valid since the condition $\mu \tau = N \omega_{c}
\tau \gg 1 $ is hold.

\subsection{Integration over the plasmon field}
\label{EffAct.Fin}

As a final step of the procedure, the action (\ref{S5}) should be
integrated over the plasmon field $\lambda$. The integration can
be performed in the quadratic approximation in the $\lambda$
fields. The corresponding propagator is determined by the second
term in Eq.(\ref{Slambda}). After that one obtains the effective
action for electrons on the partially filled Landau level which is
the main result of the paper
\begin{eqnarray}
 S_{eff} & = & - \frac{\Omega}{T} + \int d^{2} \vec r \,
 \Psi^{\dagger}(\vec r) \left [ i \omega +
\tilde \mu - {\hat {\cal H}} + i Q_{sp} \right ] \Psi(\vec r)
  -
\pi \rho \tau_{0} \int d^{2} \vec r \,  \tr Q_{sp}^{2}(\vec r)
  \nonumber \\ & - & \frac{T}{2} \int\!\!\int d^{2} \vec r \, d^{2} \vec r_{1} \, \sum
\limits_{n m k}^{\alpha} \overline \Psi^{\alpha, \sigma}_{m}(\vec
r) \Psi^{\alpha, \sigma}_{m+n}(\vec r) U_{eff}(\vec r - \vec
r_{1}) \overline \Psi^{\alpha, \sigma_{1}}_{k}(\vec r_{1})
\Psi^{\alpha, \sigma_{1}}_{k-n}(\vec r_{1}) \nonumber \\
& + & \frac{ g \omega_{c}}{2} \int \limits d^{2} \vec r \, \sum
\limits_{n}^{\alpha} \sigma \overline \Psi^{\alpha,
\sigma}_{n}(\vec r) \Psi^{\alpha, \sigma}_{n}(\vec r) \label{Sfin}
\end{eqnarray}
where we incorporate the Zeeman term into the effective action.
The Fourier transformation of the effective interaction potential
$U_{eff}(q)=U_{0}(q) / \varepsilon(q)$ is determined by the static
dielectric function $\varepsilon(q) \equiv \varepsilon(0,q)$. In
general, the low energy properties of the system concerned can be
described with the help of the retarded interaction alone (see
action (\ref{S5})). However, the description within the framework
of the effective action with the instantaneous interaction seems
to be a rather good approximation in the problem under
consideration~\cite{AG}. This is due to the fact that the
transitions between the Landau levels have a characteristic time
scale about $\omega_{c}^{-1}$ while the typical energy scale in
the effective theory is of the order of the exchange energy
$\Delta_{ex} \ll \omega_{c}$(see Sec.~\ref{Spin}).
\par The existence of the other Landau level
with the exceptions of  the partially filled $N$-th one produces
an effect on both the thermodynamic and the chemical potentials.
The thermodynamic potential $\Omega$ in action (\ref{Sfin}) can be
presented in the following way
\begin{equation}
\Omega = \Omega_{0} + \Delta \Omega \label{TP}
\end{equation}
where $\Omega_{0}$ is the thermodynamic potential of the system of
noninteracting electrons for the completely filled Landau levels
in the presence of disorder
\begin{equation}
\Omega_{0} = T \int d^{2} \vec r \, \tr \ln {\tilde G}_{0}(\vec
r,\vec r) \label{TP0}
\end{equation}
and the quantity $\Delta \Omega$ is analogous to first-order
exchange and correlation correction equivalent to the sum of ring
diagrams and contributing to the ground state energy of a clean
electron liquid~\cite{AFS}
\begin{equation}
\Delta \Omega = - \frac{T}{2} \int d^{2} \vec r \, \sum
\limits_{n} \int \frac{d^{2} { \vec q}}{(2 \pi)^{2}} \ln
\varepsilon(n,\vec q) \label{TP1}
\end{equation}
The chemical potential $\tilde \mu$ in action (\ref{Sfin}) can be
written as
\begin{equation}
\tilde \mu = \mu + \delta \mu
\end{equation}
where the shift of the chemical potential
\begin{equation}
\delta \mu = 2 \pi l^{2} T \sum \limits_{n} \int d^{2} \vec r \,
\tilde G_{0}^{n}(0,\vec r) P_{N}(0,\vec r) U_{eff}(n,\vec r)
\label{mu1}
\end{equation}
contains the corrections similar to the exchange and correlation
ones in a clean electron liquid. Here $l=1/\sqrt{m \omega_{c}}$ is
the magnetic field length. The quantity $U_{eff}(n,\vec r)$ is a
Fourier transform of $U_{0}(\vec q)/\varepsilon(n,\vec q)$ and
\begin{equation}
P_{N}(\vec r_{1},\vec r_{2}) = \sum_{k} \varphi^{*}_{N k}(\vec
r_{2}) \varphi_{N k}(\vec r_{1}) \label{pn}
\end{equation}
is the projectional operator onto the partially filled $N$-th
Landau level.
\par It should be noted that the corrections to the
thermodynamical and chemical potentials contain additional terms
except ones presented above. They are neglected in the limit of a
weak disorder $\omega_{c} \tau \gg 1$ (see Appendix A).
\par The integration over the plasmon field is performed in the
Gaussian approximation. It can be justified provided the
fluctuations of the plasmon field are small. The long and
short-range fluctuations are different physically. In the case of
large length scale $r \gg R_{c}$, the dipole transitions between
the adjacent Landau levels are induced only. The long-range
fluctuations are small if the condition $N r_{s} \gg 1$
satisfies~\cite{AG}. Physically, this condition means that the
characteristic length scale $R_{c}^{2}/a_{B}$ of the long-range
fluctuations should be much more then the cyclotron radius
$R_{c}$. The short-range fluctuations correspond to the case of
small length scale $r \ll R_{c}$. The transitions between distant
Landau levels are possible in this case. The condition $r_{s} \ll
1$ of smallness for the short-range fluctuations is just the
criterion of perturbation theory applicability to the Coulomb
interaction.

\section{Effective interaction, the thermodynamic and chemical potentials}
\label{res}

\par The results of previous section allow to find the effective
action (\ref{Sfin}) for the electrons on the partially filled
$N$-th Landau level. The main physical quantity which affects the
dynamics of the electrons is the effective electron-electron
interaction. It is completely determined by the static dielectric
function $\varepsilon(q)$. The other two interesting quantities in
the effective action (\ref{Sfin}) are the thermodynamical and
chemical potentials.

\subsection{Effective interaction}
\label{ETCP.EI}

\par The most pronounced effect of electrons on
the completely filled Landau levels is the screening of the
electron-electron interaction on the partially filled Landau
level. This screening is determined by the static dielectric
function $\varepsilon(q)$.
\par According to the equation (\ref{Pi}) for the polarization operator
$\Pi(n,q)$, the dielectric function can be obtained for arbitrary
value of the disorder parameter $\omega_{c} \tau$. However the
situation of the small broadening of Landau level due to the
disorder is the most interest from the physical point of view. In
this case the expression for the static dielectric function can be
simplified drastically
\begin{equation}
\varepsilon(q) = 1 + \frac{ 2 \pi e^{2}}{q} T \sum \limits_{n}
\pi^{n}_{0}(0,q) \, , \qquad \omega_{c} \tau \gg 1
\end{equation}
The evaluation of the static dielectric function is presented in
Appendix B. The result of evaluation can be written as
\begin{equation}
\varepsilon(q) =  1 + \frac{2}{q a_{B}} \left ( 1 - \frac{\pi}{6
\tau \omega_{c}} \right )\left (1 - {\cal J}_{0}^{2}(q R_{c})
\right ) \label{diel1}
\end{equation}
Here ${\cal J}_{0}(x)$ is the Bessel function of the first kind.
Expression (\ref{diel1}) for the static dielectric function is the
main result of the paper.
\par It is worthwhile to note that the asymptotic expressions (in
the $q R_{c} \ll 1$ and $q R_{c} \gg 1$ domains) for the static
dielectric function $\varepsilon(q)$ in a clean system ($\tau^{-1}
= 0$) was obtained earlier by Kukushkin, Meshkov and
Timofeev~\cite{KMT}. The general expression for the static
dielectric function in a clean system was derived by Aleiner and
Glazman ~\cite{AG}.
\par We mention that the asymptotic expressions for the static
dielectric function in a clean system can be obtained from a clear
physical picture~\cite{KMT, AG}. The behavior of the static
dielectric function in the region $q R_{c} \ll 1$ can be explained
by dipole transitions between the adjacent Landau levels. The
result for the static dielectric function in the region $q R_{c}
\gg 1$ is explained by the standard Thomas-Fermi screening. But
there is no clear physical picture in the case of a weak dirty
system. We have no other opportunity to obtain the dielectric
function except the derivation of the effective action for
electrons on the partially filled Landau level.
\par According to equation (\ref{diel1}) in the $q R_{c} \ll 1$
domain the static dielectric function is as follows
\begin{equation}
\varepsilon(q) = 1 + \Bigl ( 1 - \frac{\pi}{6 \omega_{c} \tau}
\Bigr ) \frac{R_{c}^{2} q}{a_{B}} \label{das1}
\end{equation}
The above result shows that the disorder suppresses the effect of
the screening. We can expect that while the disorder increases the
screening decreases. We can estimate the disorder threshold
$\tau^{*}$, i.e., the point of vanishing screening, as $\omega_{c}
\tau^{*} \sim \pi/ 6$.
\par One can obtain from equation (\ref{diel1}) the following
expression for the static dielectric function in the $q R_{c} \gg
1$ domain
\begin{equation}
\varepsilon(q) = 1 + \frac{2}{q a_{B}} \Bigl ( 1 - \frac{\pi}{6
\omega_{c} \tau} \Bigr ) \Bigl ( 1 - \frac{1+ \sin 2 q R_{c})}{\pi
q R_{c}} \Bigr ) \label{das2}
\end{equation}
The disorder suppresses the screening also in the region of large
wave vectors $q R_{c} \gg 1$.
\par Equations (\ref{diel1}) allow us to obtain
the asymptotic behavior of the effective interaction $U_{eff}(r)$
in the coordinate space. The polarization is insignificant for the
very large length scale $r \gg R_{c}^{2}/a_{B}$ and the effective
interaction coincides with the bare Coulomb interaction
\begin{equation}
U_{eff}(r) = \frac{e^{2}}{r} \left ( 1 -
\frac{R_{c}^{4}}{a_{B}^{2}r^{2}}\left [ 1 - \frac{\pi}{3
\omega_{c} \tau} \right ] \right )
\end{equation}
At the intermediate scale  $R_{c}^{2}/a_{B} \gg r \gg R_{c}$ the
polarization becomes important and the effective interaction reads
\begin{equation}
U_{eff}(r) = \frac{\omega_{c}}{\displaystyle 2 N \Bigl ( 1 -
\frac{\pi}{6 \omega_{c} \tau}\Bigr )}\ln \Bigl ( 1 +
\frac{\displaystyle R_{c}^{2}( 1 - \frac{\pi}{6 \omega_{c}
\tau})}{a_{B} r} \Bigr )
\end{equation}
We note that while disorder increases the effective interaction
tends to the bare Coulomb interaction. For the small scale $R_{c}
\gg r \gg a_{B}$ the Thomas-Fermi screening takes place and the
effective interaction has the following form
\begin{equation}
U_{eff}(r) = \frac{e^{2} a^{2}_{B} }{r^{3}} \Bigl ( 1 -
\frac{\pi}{6 \omega_{c} \tau}\Bigr )
\end{equation}
It should be emphasized that the disorder in the system affects
the electron-electron most strongly within the intermediate length
scale $R_{c}^{2}/a_{B} \gg r \gg R_{c}$. Physically, this is the
case in which dipole transitions between the adjacent Landau
levels are possible.

\subsection{The thermodynamic and chemical potentials}
\label{ETCP.TCP}

The thermodynamic and chemical potentials (\ref{TP0}-\ref{mu1})
can be evaluated in the leading orders in $1/N$. The detailed
calculations are presented in Appendix C.
\par The thermodynamic potential for the system of noninteracting
electrons in the presence of disorder for the completely filled
Landau levels is given by
\begin{equation}
\Omega_{0} = \frac{L_{x} L_{y}}{\pi l^{2}} \left [ \frac{ N ( N -
1 )}{2} \omega_{c} - \mu - \frac{\ln (2 \omega_{c} \tau ) - 1}{\pi
\tau} \right ] \label{TP01}
\end{equation}
where $L_{x}$ and $L_{y}$ are the sizes of the system.  The
first-order exchange correction to the thermodynamical potential
reads
\begin{equation}
\Delta \Omega = - \frac{L_{x} L_{y}}{\pi l^{2}} \frac{e^{2}}{\pi
l} (2 N)^{3/2} \left [ \frac{2}{3} +  \frac{2 \ln 2}{\pi
\omega_{c} \tau}\frac{1}{2 N} \right ] \label{TP11}
\end{equation}
The presence of disorder changes the dependence of $\Delta \Omega$
on the magnetic field, i.e., $N$. For the dirty system, the second
term in brackets of equation (\ref{TP11}) is proportional to
$1/N$. This is in contrast to the clean system where the
correction is much smaller and proportional to
$1/N^{2}$~\cite{AG}.
\par The shift of the chemical potential due to exchange correction
can be written as
\begin{equation}
\delta \mu = \frac{2 e^{2}}{\pi l} (2 N)^{1/2} \left [ 1 -
\frac{\ln N}{8 N} + \frac{1}{\pi \omega_{c} \tau} \frac{1}{2 N}
\right ] \label{mu2}
\end{equation}
It should be noted that $\delta \mu$ contains only the exchange
correction and does not include the correlation correction due to
normal ordering of the $\Psi^{\dagger}$ and $\Psi$ fields (see
Ref.~\cite{AG}).

\section{Spin excitations} \label{Spin}

\par In the previous section we analyzed the renormalization of the electron-electron
interaction on the partially filled $N$-th Landau level due to the
existence of the other levels. In this section we investigate the
enhancement of the $g$-factor and the simplest spin excitations at
the filling factor $\nu = 2 N + 1$.
\par The electrons on the partially filled $N$-th
Landau level at the filling factor $\nu = 2 N + 1$ possess a
maximal spin in the ground state, since the ground state does not
contain skyrmions at large $\nu$~\cite{WS}. This ground state is
obviously fully spin-polarized and described by the following wave
function $| N_{el} = N_{\Phi}, S_{z} = N_{\Phi}/2 \rangle$
 where $N_{el}$ is the number of electrons on the partially filled
$N$-th Landau level and $N_{\Phi} = L_{x} L_{y}/(2 \pi l^{2})$ is
the number of states on the Landau level. The simplest excitations
are described by the state of energy $E_{\uparrow}$ with an extra
hole and the state of energy $E_{\downarrow}$ with an extra
electron. The width of the spin gap $\Delta_{s}$ is related with
the energies of the excited states and with the energy $E_{0}$ of
the ground state~\cite{AU2,KH,AG} as follows $\Delta_{s} =
E_{\uparrow}+E_{\downarrow} - 2 E_{0}$. We can obtain that the
width of the spin gap equals $\Delta_{s} = \Delta_{ex} + g
\omega_{c}$ where the shift of the chemical potential
$\Delta_{ex}$ due to the exchange interaction~\cite{AU1,KH} is
determined by
\begin{equation}
\Delta_{ex} = 2 \pi l^{2} \int d^{2} \vec r \, U_{eff}(\vec r)
P_{N}(0,\vec r) P_{N}(\vec r,0) \label{Dex1}
\end{equation}
Using expression (\ref{pn}) for the projectional operator $P_{N}$
we can evaluate the effective $g$-factor. It is  defined as
$g_{eff} = \Delta_{s}/\omega_{c}$ and reads
\begin{eqnarray}
g_{eff}  = g +\frac{r_{s}}{\pi \sqrt{2}} \ln \Bigl [ \frac{2
\sqrt{2}}{ r_{s}}( 1 - \frac{\pi}{6 \omega_{c} \tau})\Bigr ] +
\frac{E_{h}}{\omega_{c}}\label{Geff}
\end{eqnarray}
where a ``hydrodynamic'' term is
\begin{equation}
E_{h} = \frac{\omega_{c}}{\displaystyle 2 N \Bigl ( 1 -
\frac{\pi}{6 \omega_{c} \tau} \Bigr )} \ln \Bigl [ 1 + \sqrt{2}
r_{s} N ( 1 - \frac{\pi}{6 \omega_{c} \tau} \Bigr ) \Bigr ]
\label{Hydro}
\end{equation}
The disorder in the system results in the enhancement of the
effective $g$-factor.
\par Now we are in the position to discuss the neutral excitations ---
spin waves~\cite{KH,BIE} at the filling factor $\nu = 2 N +1 $.
They are described by the following wave function
\begin{equation}
\sum \limits_{q} e^{\displaystyle i k_{x} q l^{2}} \overline
\Psi_{N,q,\downarrow} \Psi_{N,q-k_{y},\uparrow} | N_{\Phi},
\frac{N_{\Phi}}{2} \rangle \label{swf}
\end{equation}
\par Following Ref.~\cite{KH}, we should take into account three contributions.
They are the difference of the exchange self-energy of an electron
in the excited Landau level and the self-energy in the level from
which the electron removed, the direct Coulomb interaction and the
exchange energy. Then we obtain the equation which determines the
spectrum of the spin wave excitations
\begin{eqnarray}
\omega  =  g \omega_{c} +  \int \frac{d^{2} \vec q}{(2 \pi)^{2}}
\frac{U_{0}(q)}{\varepsilon(q,\omega)} \left [ L_{N}\left
(\frac{q^{2} l^{2}}{2} \right ) \right ]^{2} e^{\displaystyle -
q^{2} l^{2}/2} \Bigl ( 1 - e^{\displaystyle i { \vec k}{ \vec q}
\, l^{2}} \Bigr ) \label{SW1}
\end{eqnarray}
where $L_{N}(x)$ is the Laguerre polynomial. The dielectric
function $\varepsilon(q,\omega)$ contains the imaginary part (see
Eq.(\ref{Staticlimit})) which is of order of $1/\omega_{c} \tau$.
It results in the decay rate of the spin wave excitations.
Physically, the spin wave excitations decay due to the scattering
on impurities. We mention that the decay rate appears also in the
magnetoplasmon spectrum.
\par The energy of the spin wave excitations is much less then
$\omega_{c}$, $\omega(k) \ll \omega_{c}$. Thus we can calculate
the real $E_{SW}(k)$ and imaginary $\Gamma_{SW}(k)$ part of the
spin-wave energy separately. We put $\omega$ to zero in the right
hand side of Eq.(\ref{SW1}). Then the evaluation of equation
(\ref{SW1}) leads to a quadratic dispersion relation for the small
wave vectors $k R_{c} \ll 1$
\begin{eqnarray}
E_{SW}(k)  =  g \omega_{c} + \frac{r_{s} \omega_{c}}{\pi \sqrt{2}}
\Bigl [ 1 + \frac{r_{s}}{\sqrt{2}}  ( 1 - \frac{\pi}{ 6 \omega_{c}
\tau} ) \Bigr ]^{-1} (k R_{c})^{2} \label{SW2}
\end{eqnarray}
The disorder suppresses the effective mass of the spin wave
excitations. For the sufficiently large wave vectors $1 \ll k
R_{c} \ll R_{c}^{2}/l^{2}$, the energy of spin wave is given by
\begin{eqnarray}
E_{SW}(k) & = & \Delta_{ex} - E_{h} - \frac{r_{s} \omega_{c}}{\pi
\sqrt{2}} \Biggl [ \ln \Bigl ( 1 + \frac{ (\sqrt{2} r_{s} k
R_{c})^{-1} }{\displaystyle 1 - \frac{\pi}{ 6 \omega_{c} \tau} }
\Bigr )
 \nonumber
\\ & + &  \frac{\sin 2 k R_{c}}{2 k R_{c}} \Bigl ( 1 +
\frac{r_{s}}{\sqrt{2}} ( 1 - \frac{\pi}{6 \omega_{c} \tau}
 )\Bigr ) \Biggr ] \label{SW3}
\end{eqnarray}
\par In order to obtain the decay rate of the spin wave
excitations we take into account that the imaginary part
$\varepsilon^{''}$ of the dielectric function is small. Then we
obtain
\begin{eqnarray}
\Gamma_{SW}(k)  =   -  \int \frac{d^{2} \vec q}{(2 \pi)^{2}}
\frac{U_{0}(q)\varepsilon^{''}(q,E_{SW})}{\varepsilon_{0}^{2}(q,E_{SW})}
\left [ L_{N}\left (\frac{q^{2} l^{2}}{2} \right ) \right ]^{2}
e^{\displaystyle - q^{2} l^{2}/2} \Bigl ( 1 - e^{\displaystyle i {
\vec k}{ \vec q} \, l^{2}} \Bigr ) \label{SW4}
\end{eqnarray}
The evaluation of equation (\ref{SW4}) for the small wave vectors
$k R_{c} \ll 1$ results in
\begin{equation}
\Gamma_{SW} = - \frac{\arctan (2 \omega_{c} \tau g)}{6
\omega_{c}\tau} \frac{e^{2}}{a_{B}} (k R_{c})^{2}
\frac{1}{(\displaystyle 1 + \frac{l^{2}}{a_{B} R_{c}})^{2}}
\frac{2 - \sin 4 N}{(4 N)^{2}} \label{SW5}
\end{equation}
and for the large wave vectors $k R_{c} \gg 1$
\begin{equation}
\Gamma_{SW} = - \frac{\arctan (2 \omega_{c} \tau g_{eff})}{\pi
\omega_{c}\tau} \frac{e^{2}}{a_{B}} \Bigl [ \left
(\frac{a_{B}}{R_{c}}\right )^{2} \ln \frac{R_{c}}{a_{B}} + \frac{
\arccosh (2 k R_{c})}{2 (4 N)^{2}} \Bigr ] \label{SW6}
\end{equation}
We note that the decay rate $\Gamma_{SW}$ is the same order as the
corrections to the real part of the spin-wave energy $E_{SW}$ due
to the presence of disorder.

\section{Zero-bias anomaly}
\label{zba}

\par In this section we consider an electron tunneling into a
two-dimensional electron liquid with disorder in a weak magnetic
field. We investigate suppression of the tunneling conductivity
near zero bias, the so-called zero-bias anomaly. The properties of
an electron tunneling into the electron system are usually
described by dependence of the tunneling conductivity $G(V)$ on
bias $V$. Recently, the effective action approach for the
zero-bias problem was developed by Levitov and Shytov~\cite{LS}.
The effective action describes spreading of a tunneling electron
within the electron system in imaginary time $\zeta$.
\par Following Ref.~\cite{LS}, the action of a spreading charge
for the case of zero bias $V=0$ is determined by
\begin{equation}
S_{0}(\zeta) = 4 \int \limits_{0}^{+\infty} \frac{ d \omega}{2
\pi} \int \limits_{0}^{+\infty} \frac{ q \, d q}{2 \pi}
\frac{\sin^{2} \omega \zeta}{\omega + D q^{2}}
\frac{U_{eff}(q)}{\omega + D q^{2} + \sigma q^{2} U_{eff}(q)}
\label{s0}
\end{equation}
where $\sigma$ and $D$ are conductivity and diffusive constant of
the electron system, respectively. They relate to each other by
the Einstein's formula $\sigma = e^{2} \rho D$.
\par Using asymptotic expression (\ref{das1}) for the static
dielectric function $\varepsilon(q)$ we evaluate the action
(\ref{s0}) in the large time $\zeta \gg 1$ limit
\begin{equation}
S_{0}(\zeta) = \frac{e^{2}}{8 \pi^{2} \sigma \eta} \ln \frac{2
\zeta}{\tau_{0}} \ln \left (\frac{2 \zeta }{\tau_{0}} \beta^{4
\eta} \right ) \label{s0conc}
\end{equation}
where we introduce two dimensionless parameters
\begin{equation}
\beta =  \frac{a_{B}}{\sqrt{2} l_{el}} \qquad , \qquad\eta = \Bigl
( 1 - \frac{\pi}{6 \omega_{c} \tau} \Bigr ) \left (
\frac{R_{c}}{\sqrt{2} l_{el}} \right )^{2}
\end{equation}
with the bare elastic mean free path $l_{el} = R_{c} \omega_{c}
\tau_{0}$. According to the inequality $a_{B} \ll R_{c} \ll
l_{el}$, the parameters $\beta$ and $\eta$ are small, namely,
$\beta \ll 1$ and $\eta \ll 1$.
\par Taking into account the work
done by voltage source, we obtain the total action of the
spreading charge $S(\zeta) = S_{0}(\zeta) - 2 e V \zeta$. Then we
should find an optimal time $\zeta_{*}$ which is determined by the
minimum of the action $S(\zeta)$. The optimal time $\zeta_{*}$
plays the role of the charge accommodation time in the problem. It
can be written as
\begin{equation}
\zeta_{*} = \tau_{0} \frac{V_{0}}{2 V} \ln
\frac{V_{0}}{\displaystyle \beta^{2 \eta} V} \qquad , \qquad e
V_{0} = \Bigl ( 1 - \frac{\pi}{6 \omega_{c} \tau} \Bigr )^{-1}
\frac{1}{\pi m R_{c}^{2}} \label{V0}
\end{equation}
The theory should be self-consistent in the hydrodynamics limit,
i.e., $\zeta_{*} \geq \tau_{0}$. Hence the theory is fulfilled for
bias $V \leq V_{0}$.
\par Assuming the contribution from the barrier being just a constant at small bias,
we can write the tunneling conductivity as follows
\begin{equation}
G(V) = G_{0} \exp[ - S_{0}(\zeta_{*}) + 2 e V \zeta_{*}] \label{G}
\end{equation}
After performing evaluation we obtain the dependence of the
tunneling conductivity on small bias
\begin{equation}
G(V) = G_{0} \left (\frac{V}{V_{0}}\right )^{\alpha(V)} \qquad ,
\qquad \alpha(V) = \frac{e^{2}}{8 \pi^{2} \sigma \eta} \ln
\frac{V_{0}}{\displaystyle V \beta^{4 \eta}} \label{tunnel}
\end{equation}
\par Equation (\ref{tunnel}) shows that the screening of
the electron-electron interaction results in increasing of the
suppression of the tunneling conductivity. We mention that the
above result is valid for the bias being in the following range $V
\leq V_{0}$.
\par The expression (\ref{tunnel}) for the tunneling conductivity
contains the energy scale $e V_{0}$ which coincides with the
``hydrodynamic'' term $E_{h}$ in Eq.(\ref{Hydro}) except the
logarithm. A hydrodynamic model for the low-energy excitations of
a clean ( $\tau^{-1} = 0$ ) electron liquid in a weak magnetic
field was considered by Aleiner, Baranger and Glazman~\cite{ABG}.
They showed that the tunneling density of states exhibits a gap $2
E_{h}$ tied to the Fermi energy. Equation (\ref{Hydro}) describes
the same gap in the case of a weak disorder $\omega_{c} \tau \gg
1$. Apparently, it is disorder to be responsible for the fact that
the gap is about $0.05 \omega_{c}$ in a wide range of the applied
magnetic field~\cite{ALBS}.
\par As magnetic field increases the factor $\alpha$ increases and
becomes of order of unity. The zero-bias anomaly in the tunneling
conductivity crossovers from weak to strong. The expression
(\ref{tunnel}) shows that the factor $\alpha$ depends on bias $V$
and magnetic field. It results in the shift alone bias $V$ of the
crossover point $V_{c}$ with the change of the applied magnetic
field
\begin{equation}
V_{c} = V_{0} \exp \left ( - \frac{4 \pi
\mu}{\omega_{c}^{2} \tau_{0}} \right )
\label{cr}
\end{equation}
where $\mu$ is the chemical potential. The crossover was observed
by Ashoori et al.~\cite{ALBS} in the tunneling current from a
normal metal into a two-dimensional electrons in the presence of a
magnetic field. In the experiment the ohmic conductance was
measured as a function of temperature $T$. For low temperatures
the conductance corresponds to the zero temperature conductance
taken at $V = T/e$. The two-dimensional electrons were relatively
clean with the elastic collision time $\tau_{0} \approx 4 \cdot
10^{-12} s$. The chemical potential calculated from the electron
density was $\mu = 10 mV$. Using Eq.(\ref{cr}) the dependence of
crossover temperature on magnetic field can be written as
\begin{equation}
T_{c} = 2.9 \exp \left ( - \left [\frac{3.2}{H}\right ]^{2} \right
) \label{crexp}
\end{equation}
where temperature is measured in {\it Kelvin} while magnetic field
in {\it Tesla}. Eq.(\ref{crexp}) provides a good agreement with
the results reported in Ref.~\cite{ALBS}.

\section{Conclusion}
\label{Conc}

We considered the system of a two-dimensional electron gas in the
presence of both the disorder and the Coulomb interaction in a
weak perpendicular magnetic field. The effective low energy theory
describing electrons at the partially filled $N$-th Landau level
was derived in the case of a weak magnetic field ($ N r_{s} \gg 1
$) and a weak interaction ($r_{s} \ll 1$). The modified
electron-electron interaction for electrons on the partially
filled Landau level takes into account the screening from the
other electrons on the occupied Landau levels. We also presented
the exchange corrections to the thermodynamical and chemical
potentials in the presence of disorder.
\par The theory proposed allows us to take into account the effects
of disorder in the problems connected with the behavior of a
two-dimensional electron gas in the weak magnetic field. How
disorder affects the formation of stripes, bubble
phase~\cite{TBP}, tunneling density of states, spin excitations,
 tunneling conductivity and so on can be investigated.
\par We discussed the effect of disorder on the exchange enhancement of the $g$-factor and
 the simplest spin excitations on the partially filled Landau level.
We obtained an additional dependence of the effective $g$-factor
as a function of the magnetic field, the suppression of the
effective mass  and the decay rate of the spin wave excitations.
\par We also investigated an electron tunneling into a
two-dimensional electron liquid with a weak disorder in a weak
magnetic field. We obtained the enhancement of the gap in the
tunneling density of states and nonlinear dependence of tunneling
conductivity on the applied bias.

\section{Acknowledge}
\label{Ack}

Author is grateful to M.A. Baranov, M.V. Feigel'man, and A.S.
Iosselevich for useful discussions. Author thanks M.A. Baranov for
some critical remarks during the manuscript preparation.

\renewcommand{\theequation}{A.\arabic{equation}}
\setcounter{equation}{0}
\section{Appendix A}
\label{A}

In this appendix the calculation of the term $S_{\mu}$ in action
(\ref{S5}) is presented. This term appears after performing the
integration over the longitudinal fluctuations and equals
\begin{equation}
S_{\mu} = \frac{1}{2} \langle \left ( S_{2}[\overline \Psi,\Psi,
\delta P,\lambda]\right )^{2} \rangle_{\delta P}
\end{equation}
where $\langle \cdots \rangle_{\delta P}$ means the average with
the propagator of the $\delta P$ field (\ref{Pcor}). One can
obtain
\begin{equation}
S_{\mu} = \delta S_{1} + \delta S_{2} + \delta S_{3} \label{A.mu}
\end{equation}
where
\begin{equation}
\delta S_{1} = T^{2} \int\!\!\int d^{2} \vec r_{1} \, d^{2} \vec r
\,_{2} \left ( \Psi^{\dagger}  \hat \lambda  \tilde G_{0} \hat
\lambda \Psi - 2 \tr  \hat \lambda \tilde G_{0} \hat \lambda
G_{0}\right )
\end{equation}
\begin{eqnarray}
\delta S_{2}   =   -  T^{2} \int d^{2} \vec r_{1} \,\!\cdots\!\int
d^{2} \vec r \,_{4} \sum \limits^{\alpha \beta}_{k,n,m}
\lambda^{\beta}_{m}(\vec r_{1}) D_{n m}^{\alpha \alpha}(k;\vec
r_{3},\vec r_{4}) \pi_{0}^{n}(m;\vec r_{1},\vec r_{2}) L_{n,
n+m}^{\alpha \beta,\alpha \alpha}(\vec r_{2},\vec r_{4})
\end{eqnarray}
\begin{eqnarray}
\delta S_{3}  =  \frac{T^{2}}{2} \int d^{2} \vec r_{1}
\,\!\cdots\!\int d^{2} \vec r \,_{4} \sum \limits^{\alpha \beta
\gamma \delta}_{k,l,n,m}
D_{n m}^{\alpha \beta}(k;\vec r_{1},\vec r_{2}) D_{n m}^{\gamma
\delta}(l;\vec r_{3},\vec r_{4}) L_{n, n+m}^{\alpha \beta, \gamma
\delta }(\vec r_{2},\vec r_{4})
\end{eqnarray}
Here
\begin{eqnarray}
& &  D_{n m}^{\alpha \beta}(k;\vec r_{1},\vec r_{2})  \nonumber \\
& = & \left ( \overline \Psi^{\alpha}_{k}(\vec r_{1})
\Psi^{\beta}_{n+m}(\vec r_{2}) - 2 \delta_{k,n+m}\delta^{\alpha
\beta} G_{0}^{n+m}(\vec r_{2},\vec r_{1}) \right )
\lambda^{\alpha}_{n-k}(\vec r_{1}) \tilde G_{0}^{n}(\vec
r_{1},\vec r_{2})  \nonumber \\ & + & \left( \overline
\Psi^{\beta}_{n}(\vec r_{1}) \Psi^{\alpha}_{k}(\vec r_{2}) - 2
\delta_{k,n}\delta^{\alpha \beta} G_{0}^{n}(\vec r_{2},\vec r_{1})
\right ) \lambda^{\beta}_{k-n-m}(\vec r_{1}) \tilde
G_{0}^{n+m}(\vec r_{1},\vec r_{2})
\end{eqnarray}
and  $L_{m_{1} m_{2}}^{\alpha \beta,\gamma \delta}$ is the
propagator of the longitudinal fluctuations (\ref{Pcor}).
\par Performing the integration over the plasmon field, one obtains from
the $S_{\mu}$ term
\begin{equation}
S_{\mu} \to \frac{\Delta \Omega_{1} + \Delta \Omega_{2} + \Delta
\Omega_{3} }{T} + (\delta \mu_{1} + \delta \mu_{2} + \delta
\mu_{3}) \int d^{2} \vec r \, \Psi^{\dagger}(\vec r) \Psi(\vec r)
\end{equation}
where second-order corrections to the thermodynamic potential
 are given by
\begin{eqnarray}
\frac{\Delta \Omega_{1}}{ L_{x} L_{y}} & = & - T^{2} \sum
\limits_{n m} \int d^{2} \vec r \, G_{0}^{n}(0,\vec r) \tilde
G_{0}^{m}(\vec r,0)
U_{eff}(m-n,\vec r) \\
\frac{\Delta \Omega_{2}}{ L_{x} L_{y}} & = & 2 T^{2} \sum
\limits_{n m} \int d^{2} \vec r \, d^{2} \vec r_{1} \, d^{2} \vec
r_{2}\,
\left ( G_{0}^{n}(0,\vec r) \tilde G_{0}^{n+m}(\vec
r,0) + \tilde G_{0}^{n}(\vec r,0) G_{0}^{n+m}(0,\vec r) \right )
\nonumber \\
& \times & U_{eff}(m,\vec r_{1}) \pi_{0}^{n}(m,\vec r_{1}-\vec
r_{2})
L_{n,n+m}^{\alpha \alpha,\alpha \alpha}(\vec r_{2} - \vec r) \nonumber \\
\frac{\Delta \Omega_{3}}{ L_{x} L_{y}} & = & 2 T^{2} \sum
\limits_{n m} \int d^{2} \vec r \, d^{2} \vec r_{1} \, d^{2} \vec
r_{2}\,
\left ( G_{0}^{n}(0,\vec r) \tilde G_{0}^{n+m}(\vec
r,0) + G_{0}^{n+m}(0,\vec r) \tilde G_{0}^{n}(\vec r,0) \right )
\nonumber \\
& \times & \left ( G_{0}^{n}(0,\vec r_{1}) \tilde G_{0}^{n+m}(\vec
r_{1},0) + G_{0}^{n+m}(0,\vec r_{1}) \tilde G_{0}^{n}(\vec
r_{1},0) \right )
U_{eff}(m,\vec r_{2}) L_{n,n+m}^{\alpha \alpha,\alpha \alpha}(\vec
r - \vec r_{1} +\vec r_{2}) \nonumber \label{domega}
\end{eqnarray}
The above corrections are negligible in the parameter $N^{-1}$
compared with the correction determined by equation (\ref{TP1}).
\par The corrections to the chemical potential are such as
\begin{eqnarray}
\frac{\delta \mu_{1}}{2 \pi l^{2}} & = & T \sum \limits_{m} \int
d^{2} \vec r \, P_{N}(0,\vec r) \tilde G_{0}^{m}(\vec r,0)
U_{eff}(m,\vec
r)  \\
\frac{\delta \mu_{2}}{2 \pi l^{2}} & = & -4  T \sum \limits_{m}
\int d^{2} \vec r \, d^{2} \vec r_{1} \, d^{2} \vec r_{2}\,
P_{N}(0,\vec r) \tilde G_{0}^{m}(\vec r,0)
U_{eff}(m,\vec r_{1}) \pi_{0}^{0}(m,\vec r_{1}-\vec r_{2})
L_{0,m}^{\alpha \alpha,\alpha \alpha}(\vec r_{2} - \vec r)
\nonumber \\
\frac{\delta \mu_{3}}{2 \pi l^{2}}  & = & - 8 T^{2} \sum
\limits_{m} \int d^{2} \vec r \, d^{2} \vec r_{1} \, d^{2} \vec
r_{2}\, P_{N}(0,\vec r) \tilde G_{0}^{m}(\vec r,0)
U_{eff}(m,\vec r_{1}) L_{0,m}^{\alpha \alpha,\alpha \alpha}(\vec r
- \vec r_{1}+\vec r_{2})  \nonumber \\ & \times & \left (
G_{0}^{0}(0,\vec r_{1}) \tilde G_{0}^{m}(\vec r_{1},0) +
G_{0}^{m}(0,\vec r_{1}) \tilde G_{0}^{0}(\vec r_{1},0) \right )
\nonumber \label{dmu}
\end{eqnarray}
The second and third corrections are negligible in the parameter
$N^{-1}$ compared with the first term. Hence the shift of the
chemical potential $\delta \mu$ is determined mainly by the first
correction $\delta \mu_{1}$.

\renewcommand{\theequation}{B.\arabic{equation}}
\setcounter{equation}{0}
\section{Appendix B}
\label{B}

In this appendix the evaluation of the polarization operator
$\Pi(\omega_{n},q)$ is presented. The condition $\omega_{c} \tau
\gg 1$ is assumed to be hold. Then
\begin{equation}
\Pi(\omega_{n},q) = T \sum \limits_{m} \pi^{m}(\omega_{n},q)
\end{equation}
The calculation of the polarization operator $\Pi(\omega_{n},q)$
is
 analogous to that given in Ref.~\cite{AG}. The wave vectors $q \ll R_{c}/l^{2}$
 are considered.
\par Using equation (\ref{Pi}), one immediately obtains
\begin{equation}
\Pi(\zeta_{n}, Q)  =  \frac{2 m}{\pi} \sum \limits_{j=1}^{\infty}
\frac{{\cal J}_{j}^{2}(Q)}{j^{2}+ \zeta_{n}} \Biggl [ j^{2} -
\frac{1}{\pi \tau \omega_{c}} {\mathcal L}_{j}(\zeta_{n}) \Bigr )
\Biggr ] \label{piser}
\end{equation}
where
\begin{eqnarray}
{\mathcal L}_{j}(\zeta_{n}) = \frac{j^2(j^2+3
\zeta_{n}^{2})}{\zeta_{n}^{2}(j^{2}+\zeta_{n}^{2})} \ln
\frac{\sinh \pi \zeta_{n}}{\pi \zeta_{n}} + 4 \frac{j
\zeta_{n}}{j^{2}+\zeta_{n}^{2}} \arctan\frac{\zeta_{n}}{j} +
\frac{j^{2}-\zeta_{n}^{2}}{j^{2}+\zeta_{n}^{2}} \ln (1+2 \tau
\omega_{c} \zeta_{n}) \label{L}
\end{eqnarray}
 Here two parameters $\zeta_{n} = \omega/\omega_{c}$ ($\omega =
2 \pi T n$) and $Q= q R_{c}$ are introduced. The transformation of
a series (\ref{piser}) into the integral form yields the
asymptotic form of the polarization operator in the different
regimes. In the static limit $\zeta_{n} \ll 1$
\begin{equation}
\Pi(\zeta_{n},Q)  = \frac{m}{\pi} \left ( \left (1-\frac{\pi}{6
\tau \omega_{c}} \right )( 1 - {\cal J}_{0}^{2}(Q)) + \frac{\ln
(1+2 \omega_{c}\tau \zeta_{n})}{2 \pi \omega_{c}\tau} \xi(Q)+
{\cal O}(\zeta_{n}^{2}) \right ) \label{Staticlimit}
\end{equation}
where the function $\xi(x)$ is defined as
\begin{equation}
\xi(x) = \int \limits_{0}^{\pi} \frac{d y}{\pi} {\cal J}_{0}(2 x
\sin\frac{y}{2}) \left [ (y-\pi)^{2}- \frac{\pi^{2}}{3}\right ]
\end{equation}
and its asymptotic form is given by
\begin{equation}
\xi(x) = \left \{
\begin{array}{lcr}
x^{2} & , & x \ll 1 \\
\displaystyle \frac{\pi}{3 x}(2 - \sin 2 x) & , & x \gg 1
\end{array}
\right .
\end{equation}
In the hydrodynamic limit $q R_{c} \ll 1$ we obtain
\begin{eqnarray}
\Pi(\zeta_{n},Q) & = & \frac{m}{2 \pi} \frac{Q^{2}}{1 +
\zeta_{n}^{2}} \Bigl [ 1 - \frac{1}{\pi \tau \omega_{c}} {\cal
L}_{1}(\zeta_{n})\Bigr ] \label{Hydrolimit}
\end{eqnarray}

\renewcommand{\theequation}{C.\arabic{equation}}
\setcounter{equation}{0}
\section{Appendix C}
\label{C}

In this appendix the evaluation of the corrections to the
thermodynamic and chemical potentials are presented.

\subsection{Thermodynamic potential correction}

Using equation (\ref{TP1}), one can separate the thermodynamic
potential correction into the exchange and correlation ones as
follows
\begin{equation}
\Delta \Omega = \Delta \Omega_{ex} + \Delta \Omega_{c}
\end{equation}
\begin{equation}
\frac{\Delta \Omega_{ex}}{L_{x} L_{y}} = \frac{T}{2} \sum
\limits_{n} \int \frac{ d^{2} \vec q}{(2 \pi)^{2}} U_{0}(q)
\Pi(n,q)
\end{equation}
\begin{equation}
\frac{\Delta \Omega_{c}}{L_{x} L_{y}} = -  \frac{T}{2}\sum
\limits_{n} \int \frac{ d^{2} \vec q}{(2 \pi)^{2}} \int
\limits_{0}^{1} d \alpha  \frac{ \alpha U_{0}^{2} (q) \Pi^{2}(n,q)
}{1 + \alpha U_{0}(q) \Pi(n,q)}
\end{equation}
The exchange correction gives the leading contribution~\cite{AG}
and can be written in the following way
\begin{eqnarray}
\frac{\Delta \Omega_{ex}}{L_{x} L_{y}} & = & - \frac{e^{2}}{2 \pi
l^{3}}  \sum \limits_{m \neq N} \int \limits_{0}^{\infty}d x\,
e^{-x^{2}/2} L_{N}^{1}(\frac{x^{2}}{2}) L_{m}(\frac{x^{2}}{2})
\left ( \Theta(N-m)  + \frac{1}{\pi \omega_{c} \tau}
 \frac{1}{m-N}\right )
\end{eqnarray}
where $L_{n}^{m}$ stands for the Laguerre polynomials. The above
equation in the case $N \gg 1$ goes over into equation
(\ref{TP11}).

\subsection{Chemical potential correction}

Using equation (\ref{mu1}), one can separate the chemical
potential correction onto exchange and correlation ones,
respectively,
\begin{equation}
\delta \mu = \delta \mu_{ex} + \delta \mu_{c}
\end{equation}
\begin{equation}
\delta \mu_{ex} = 2 \pi l^{2} T \sum \limits_{n} \int d^{2} \vec r
\, U_{0}(r) P_{N}(0,\vec r) \tilde G^{n}_{0}(\vec r,0)
\end{equation}
\begin{equation}
\delta \mu_{c} = - 2 \pi l^{2} T \sum \limits_{n} \int \frac{
d^{2} \vec q}{(2 \pi)^{2}} P_{N}(q) \tilde G^{n}_{0}(q)\frac{
U_{0}^{2}(q) \Pi(n,q) }{1 +  U_{0}(q) \Pi(n,q)}
\end{equation}
The exchange correction gives the leading contribution~\cite{AG}
and can be written in the following way
\begin{eqnarray}
\delta \mu_{ex} & = & - \frac{e^{2}}{l} \sum \limits_{m \neq N}
\int \limits_{0}^{\infty}d x\, e^{-x^{2}/2} L_{N}(\frac{x^{2}}{2})
L_{m}(\frac{x^{2}}{2})
\left ( \Theta(N-m)  + \frac{1}{2 \pi \omega_{c} \tau}
 \frac{1}{m-N}\right )
\end{eqnarray}
The above equation in the case $N \gg 1$ leads to
\begin{equation}
\delta \mu_{ex} =  \frac{ 2 e^{2}}{\pi l^{2}}(2 N)^{1/2} \Biggl [
1 - \frac{\ln N}{8 N} + \frac{1}{4 \pi \omega_{c} \tau} \frac{1}{2
N}
 \Bigl ( \int \limits_{1}^{\infty} \frac{d t}{t}
\ln (1- e^{-t}) + \int \limits_{0}^{1} \frac{d t}{t} \ln \frac{1-
e^{-t}}{t} - \frac{\pi^{2}}{3} \Bigr ) \Biggr ]
\end{equation}



\end{document}